\documentstyle[twoside,fleqn,npb,epsfig]{article}
\newcommand{\AmS}{{\protect\the\textfont2
  A\kern-.1667em\lower.5ex\hbox{M}\kern-.125emS}}

\def    \np     #1#2#3{{\it Nucl. Phys.} {\bf #1}(19#2)#3}

\def    \pr     #1#2#3{{\it Phys. Rev.} {\bf #1}(19#2)#3}

\def    \zp     #1#2#3{{\it Z. Phys.} {\bf #1}(19#2)#3}
\def    \ibid   #1#2#3{{\it ibid.} {\bf #1}(19#2)#3}
\def    \jhep   #1#2#3{{\it JHEP} {\bf #1}(19#2)#3}
\def    \hepph  #1 {{\tt hep-ph/#1}}
\def    \hepex  #1 {{\tt hep-ex/#1}}
\def\beq{\begin{equation}}
\def\eeq{\end{equation}}
\def\beqn{\begin{eqnarray}}
\def\eeqn{\end{eqnarray}}
\def\sss{\scriptscriptstyle}
\def\mq{m_{\sss Q}}
\def\mqt{m_{\sss Q}^2}
\def\bQ{\overline{Q}}
\def\mur{\mu_{\sss R}}
\def\muf{\mu_{\sss F}}
\def\murp{\mu_{\sss R}^\prime}
\def\mufp{\mu_{\sss F}^\prime}
\def\mug{\mu_{\sss \gamma}}
\def\pt{p_{\sss T}}
\def\ptt{p_{\sss T}^2}
\def\aem{\alpha_{em}}
\def\as{\alpha_{\sss S}}
\def\asbare{\alpha_{\sss S}^{bare}}
\def\asren{\alpha_{\sss S}^{ren}}
\def\MSbar{{\overline{\rm MS}}}
\def\ep{\epsilon}
\def\epb{\bar{\epsilon}}
\def\CA{C_{\sss A}}
\def\TF{T_{\sss F}}
\def\Nlf{N_{lf}}
\def\HQ{H_{\sss Q}}

\begin{document}

\twocolumn[
\vskip -3cm
~

\flushright{
        \begin{minipage}{2.9cm}
        CERN-TH/99-157\hfill \\
        hep-ph/9905545\hfill \\
        \end{minipage}        }

\vskip 5.0em
\flushleft{\Large Heavy quark photoproduction}~\footnotemark
\vskip\baselineskip
\flushleft{Stefano Frixione$^{\rm a}$}
\vskip\baselineskip
\flushleft{$^{\rm a}$CERN, TH division, Geneva, Switzerland}

\begin{center}
\begin{minipage}{160mm}
                \parindent=10pt
{\small 
I present a review of selected topics in the computation of 
heavy flavour cross sections in photon-hadron collisions.
}
                \par
                \end{minipage}
                \vskip 2pc \par
\end{center}
]
\footnotetext{Talk given at DIS99, 19-23 April 1999, Zeuthen, DE.}


\section{Computation of cross sections at fixed order in QCD}

A heavy quark $Q$ is by definition a quark whose mass $\mq$ is much 
larger than $\Lambda_{\sss QCD}$. Since $\mq$ is different from zero,
it cuts off the singularities due to the collinear emission of
gluons from the heavy quark lines, thus allowing the definition
of open-heavy-quark cross sections, that are infrared-finite
order by order in perturbative QCD. The hard scale of the 
production process is of the order of $\mq$. Therefore, one
expects the perturbative predictions to be very reliable in the
case of top, under control in the case of bottom, and affected
by large uncertainties in the case of charm. This is more so, since
the smaller the mass, the bigger the impact that effects of 
non-perturbative origin will have on the predictions of physical 
observables. The cross section for the production of $Q\bQ$ pairs 
in on-shell photon-hadron reactions is usually written as the sum of 
two terms:
\beqn
&&\!\!\!\!\!\!\sigma_{\gamma H}=
f_j^{(H)}(\muf)\otimes\hat{\sigma}_{\gamma j}(\mur,\muf,\mug)
\label{factth}
\\&&\phantom{\!\!\!\!\!\!\sigma_{\gamma H}}
+f_i^{(\gamma)}(\mug)\otimes f_j^{(H)}(\mufp)
\otimes\hat{\sigma}_{ij}(\murp,\mufp,\mug),
\nonumber
\eeqn
denoted as pointlike and hadronic components respectively. Neither
of the two is well defined beyond LO in perturbation theory. 
This fact is formally taken into account in eq.~(\ref{factth}) 
by the dependence upon the scale $\mug$, which is related to
the subtraction of singularities due to the collinear splitting
of the incoming photon. The pointlike and hadronic components
are not constant with respect to a variation of $\mug$, 
while their sum is (up to higher order terms in perturbation 
theory). On the other hand, the pointlike and hadronic components 
are {\it separately} constant with respect to a variation of all the other 
mass scales which enter eq.~(\ref{factth}). The contribution of the 
hadronic component will be larger (relative to the pointlike component) 
the smaller $\mq$. However, it has to be remarked that kinematical 
configurations exist where the hadronic component is basically negligible:
typical examples are the high-$\pt$ region of the emitted heavy quarks,
and small center-of-mass energy collisions (like those at
fixed-target experiments).  This is due to the fact that photon 
densities $f^{(\gamma)}_i(x)$ are vanishing faster than hadron 
densities for $x\to 1$.

The task for theorists is to calculate the partonic cross sections
$\hat{\sigma}_{\gamma i}$ and $\hat{\sigma}_{ij}$ at the highest 
possible order in perturbation theory. In 
ref.~\cite{calcpnt} (for the pointlike component) and in 
ref.~\cite{calchad} (for the hadronic component)
results accurate to NLO ($\aem\as^2$ and $\as^3$ respectively)
have been obtained. At that time, predictions were only available
for total cross sections and single-inclusive spectra. Later, 
computations were performed~\cite{fullyexcl}, which allowed
the definition of cross sections fully exclusive in terms of the
variables of $Q$, $\bQ$, and of the accompanying jet. The matrix 
elements relevant for $Q\bQ$ production are mostly convenient
calculated in a modified $\MSbar$ scheme~\cite{MSbarmod}, in which 
the renormalized $\as$ is defined through the following equation
(up to terms of order $\as^3$):
\beqn
&&\!\!\!\!\!\!\!\!\!\asbare=\mu^{2\ep}\asren\left[1-\frac{\asren}{4\pi\epb}
\frac{11\CA-4\TF\Nlf}{3}\right]
\nonumber \\&&\phantom{\!\!\!\!\!\!\!\!\!\asbare}\times
\left[1+\frac{\asbare}{4\pi\epb}\frac{4\TF m_{\sss Q}^{-2\ep}}{3}\right],
\label{alphas}
\eeqn
where $\Nlf$ is the number of flavours lighter than $Q$ (i.e., 3 for
charm, 4 for bottom and 5 for top). The first term in square brackets
in eq.~(\ref{alphas}) is the same that appears in the standard
$\MSbar$ scheme; thus, light flavours are dealt with like in standard
$\MSbar$.  The second term in square brackets is on the other hand
peculiar of heavy flavour production, and is designed in such a way
that, for momenta much smaller than $\mq$, it cancels exactly the
contributions of heavy quark divergent loops.  Since this second term
is only expressed in terms of $\asbare$, it follows that the
renormalization group equation one derives is equal to that of the standard
$\MSbar$ {\it with $\Nlf$ flavours}. This is consistent with the fact
that no contribution to the cross section comes from diagrams where a
heavy quark is present in the initial state.

The cross sections computed in this way are expected to give sensible
predictions for momenta of the order of $\mq$. At the borders of the
phase space, or for high energies, some logarithmic terms may grow
large and spoil the convergence of the perturbative series. Example of
large logarithms are: $\log(S/m_{\sss Q}^2)$ (small-$x$ effects: they
are important when the center-of-mass energy of the colliding
particles is large); $\log(p_{\sss T}^{Q\bQ}/\mq)$ (due to the
emission of soft gluons, they are important at the edges of the phase
space or at the threshold); $\log(\pt/\mq)$ (due to the emission of
collinear gluons from the heavy quark line, they are important when
$\pt\gg\mq$). It has been shown that the impact of small-$x$ effects
for charm and bottom production at present colliders is moderate.
Also, soft gluon resummation gives small enhancements with respect to
NLO results for total cross sections in the case of top production at
the Tevatron and of bottom production at HERA-B. In the following, I
will concentrate on collinear logarithms, which are phenomenologically
relevant for charm photoproduction at HERA.

\section{The large transverse momentum region}

If $\pt\gg\mq$, mass terms in the matrix elements are suppressed by
powers of $\mq/\pt$. Therefore, when computing matrix elements one
is entitled to treat the heavy quark as massless~\cite{CG,KK} and use 
the standard $\MSbar$ scheme (this is the reason why this kind of 
calculations are often - incorrectly - referred to as ``massless''). 
This implies that it is mandatory to consider 
diagrams where the heavy quark is also present in the initial
state. Furthermore, the final state $Qg$ collinear singularities
are no longer cut off by the quark mass. The subtraction of these
additional singularities is performed within the framework of the
factorization theorem; one writes the cross section for the production
of a heavy quark (in any kind of hard collisions) as follows:
\beq
d\sigma_{\sss Q}=d\sigma_i(\mu)\otimes D_{\sss Q}^{(i)}(\mu),
\label{sigmares}
\eeq
where $i$ can be a gluon, a light quark or the heavy quark $Q$, and
$d\sigma_i$ is the cross section for the production of the flavour $i$.
$D_{\sss Q}^{(i)}$ are the so-called perturbative fragmentation
functions (PFF), whose bare form contain a divergent term which
cancels the final-state collinear singularities mentioned above. As far
as the $\mu$ dependence is concerned, the PFFs obey the usual
Altarelli-Parisi equations.  However, at variance with the case of
parton densities, the PFFs are fully calculable in perturbation
theory, for scales $\mu=\mu_0$ of the order of the heavy quark mass
(this prevents $D_{\sss Q}^{(i)}(\mu_0)$ from containing large
logarithms).  Such a calculation has been performed in
ref.~\cite{MeleNason}, up to terms of order $\as^2$. At this order,
$D_{\sss Q}^{({\sss Q})}$ and $D_{\sss Q}^{(g)}$ are non-vanishing, while
$D_{\sss Q}^{(q)}$ is zero. The main feature of eq.~(\ref{sigmares})
is the following: the scale $\mu$ is of the order of $\pt$, which is
the hard scale of the process. Since in the computation of matrix
elements the mass of the heavy quark has been set to zero, $d\sigma_i$
does not contain any large logarithm of $\pt/\mq$.  All the mass
effects are included in $D_{\sss Q}^{(i)}$; however, here the large
logarithms are resummed by the Altarelli-Parisi equations, in the
evolution from the scale $\mu_0$ to the scale $\mu$.

If the data are presented in terms of heavy-flavoured hadrons $\HQ$,
theoretical predictions must be given for such quantities;
these are calculated starting from the heavy-quark cross section
given in eq.~(\ref{sigmares}), and convoluting it with a non-perturbative
fragmentation function (NPFF), that accounts for the hadronization
process $Q\to \HQ$:
\beq
d\sigma_{\sss \HQ}=d\sigma_{\sss Q}\otimes D_{\sss \HQ}^{({\sss Q})}
\label{sigmameson}
\eeq
(of course, the convolution with a NPFF can be also performed starting
from a fixed-order QCD cross section, like the one in
eq.~(\ref{factth})).  The NPFFs can not be calculated in perturbation
theory; however, they are supposed to be universal. Therefore, they
can be determined by a fit to experimental data -- typically, from
$e^+e^-$ collisions -- and eventually used to obtain theoretical 
predictions for various production processes.  Recent fits are presented 
in refs.~\cite{NO,CGtwo}; it turns out that the NPFF is much harder than
in previous determinations based on LO QCD analysis.  A different
procedure has also been proposed~\cite{KK}.  Instead of considering
PFFs and NPFFs, one can introduce a fragmentation function which
describes at the same time the hard and soft physics: $D_{\sss
  \HQ}^{(i)}=D_{\sss Q}^{(i)}\otimes D_{\sss \HQ}^{({\sss Q})}$. The idea is
then to parametrize $D_{\sss \HQ}^{(i)}$, fit it to the data, and then
evolve it to the appropriate scale with Altarelli-Parisi equations.
However, since the NPFFs considered in practice do not depend upon a
mass scale, there is no difference in evolving only the PFF or the
convolution of the PFF and the NPFF (this can be trivially shown by
writing the Altarelli-Parisi equations in Mellin space). Therefore,
ref.~\cite{CG} and~\cite{KK} {\it only} differ in the choice of the
initial conditions for the evolution of the fragmentation function.
While in ref.~\cite{CG} the $\as$ terms of $D_{\sss Q}^{(i)}(\mu_0)$
are taken into account, they are not included in ref.~\cite{KK}.
Thus, ref.~\cite{KK} only includes mass effects through a fitting to
data, and does not exploit all the perturbative results available.

\section{Matching large and small momentum results}

As discussed in the previous sections, if $\pt$ is of
the order of $\mq$, then fixed-order QCD computations have to be used;
on the other hand, if $\pt$ is much larger than $\mq$, resummed 
calculations are mandatory. What is missing, unfortunately, is
a quantitative definition of what ``much larger'' means. HERA data
are compared to resummed calculations for $p_{\sss T}(D^*)$ down
to $2\div 3$~GeV, which means $\pt(c)$ on average smaller than 
4~GeV (here and in what follows, I take 
$p_{\sss T}(D^*)=\langle z\rangle p_{\sss T}(c)$, 
where $\langle z\rangle$ is the average $z$ obtained 
using a Peterson fragmentation function with $\ep=0.03$). 
The ratio 4~GeV$/m_c$ does not seem to be large enough. 
Any agreement between data and resummed cross sections in
this $\pt$ region should therefore be regarded as incidental.
However, it is clear that there is an intermediate region in $\pt$ 
where both fixed-order and resummed calculations are almost equally 
reliable (or, if one adopts a pessimistic point of view, neither of 
the two is meaningful). Therefore, it is conceivable to have a 
matched calculation, which returns a fixed-order result at low
$\pt$, a resummed result at large $\pt$, and smoothly connects
the two approaches in the intermediate region. Unfortunately, such
a calculation does not exist for photoproduction. However, results
are available for hadroproduction~\cite{CGN,OST}. It is instructive
to see how this matter is dealt with in ref.~\cite{CGN}, since
there is no conceptual difference with respect to the case of
photoproduction (to recover the latter, simply substitute
$\as^n$ with $\aem\as^{n-1}$ in the following equations).
The fixed (NL)-order and (NLL) resummed cross sections are written as
($\mu$ is a scale of the order of $\pt$):
\beqn
&&\left(\frac{d\sigma}{d\ptt}\right)_{\sss NLO}=
\as^2 A + \as^3 B + {\cal O}(\as^4),
\label{sigFO}
\\
&&\left(\frac{d\sigma}{d\ptt}\right)_{\sss RES}=
\as^2\sum_{n=0}^\infty a_n\left(\as\log\frac{\mu}{\mq}\right)^n
\nonumber \\&&\phantom{\left(\frac{d\sigma}{d\ptt}\right)_{\sss RES}}
+\as^3\sum_{n=0}^\infty b_n\left(\as\log\frac{\mu}{\mq}\right)^n
\nonumber \\&&\phantom{\left(\frac{d\sigma}{d\ptt}\right)_{\sss RES}}
+{\cal O}\left(\as^4\left(\as\log\frac{\mu}{\mq}\right)^n\right).
\label{sigRES}
\eeqn
The coefficients $A$ and $B$ contain logarithms of $\pt/\mq$. 
Eq.~(\ref{sigRES}) is defined up to power-suppressed terms, not 
explicitly indicated. We can now define the matched cross section:
\beqn
&&\!\!\!\!\!\!\!\!\!\!\!\frac{d\sigma}{d\ptt}=
\as^2 A + \as^3 B 
+\Bigg\{\as^2\sum_{n=2}^\infty a_n\left(\as\log\frac{\mu}{\mq}\right)^n
\nonumber \\&&\phantom{\!\!\!\!\!\!\!\!\!\!\!\frac{d\sigma}{d\ptt}}
+\as^3\sum_{n=1}^\infty b_n\left(\as\log\frac{\mu}{\mq}\right)^n\Bigg\}
\,G(\mq,\pt),
\label{sigmatched}
\eeqn
where the function $G$ is arbitrary to a large extent, but must
approach 1 when $\mq/\pt\to 0$. If $\pt\simeq\mq$, the quantity in
curly brackets vanishes, and the matched cross section approaches
eq.~(\ref{sigFO}). If $\pt\gg\mq$, the matched cross section
approaches eq.~(\ref{sigRES}), since logarithmic terms in $A$ and $B$
dominate (by construction, these terms are exactly those missing in
the sums of eq.~(\ref{sigmatched}) with respect to the sums of
eq.~(\ref{sigRES})). Notice that this is true regardless of the form
of $G$, provided that it satisfies the condition given above. One
could therefore choose $G\equiv 1$. However, as shown in
ref.~\cite{CGN}, this choice can lead to instabilities in the
numerical computations. In ref.~\cite{CGN}, the following form has
been adopted:
\beq
G(\mq,\pt)=\frac{\ptt}{\ptt+\beta^2\mqt},
\eeq
where $\beta$ is a suitable number. From numerical studies performed
for bottom production at the Tevatron, it turns out that a sensible
choice is $\beta=5$. This means that, in the matched cross section,
the whole tower of logarithms (except those with coefficients $a_0$,
$a_1$ and $b_0$) is suppressed by a factor 0.5 for $\pt=5\mq$,
and is almost fully included ($G=0.9$) for $\pt=15\mq$.
If we assume that these conclusions can be also applied to charm
production at HERA, we conservatively conclude that a resummed cross 
section can be safely compared to the data starting from a $\pt(c)$ 
of the order of $22$~GeV (here, $m_c=1.5$~GeV), that is for $\pt(D^*)$
of the order of $16$~GeV. Being less conservative, we can say that 
the {\it lowest} $\pt(c)$ value for which a comparison between the data 
and a resummed cross section is meaningful is of the order of $8$~GeV 
($\pt(D^*)$ of the order of $5\div 6$~GeV).

\section{Conclusions}

I presented a review of results relevant for the production of heavy
flavours in high-energy photon-hadron collisions. At low and moderate
transverse momenta, where the bulk of the total cross section comes
from, fixed-order QCD computations (available to NLO accuracy) are
appropriate. HERA data on charm production have renewed interest in
the resummation of large logarithms of the transverse momentum. A
brief discussion on the techniques designed to perform this
resummation has been given. Finally, it has been argued that most of
the currently available data for $D^*$ production at HERA have too 
small a $\pt$ to be safely compared with a resummed calculation. A
definite statement on this issue will however only come when a matched
calculation will be available, which combines the results of fixed-order 
and resummed computations.

\noindent
{\bf Acknowledgements}: I would like to thank Matteo Cacciari for
useful discussions.

\end{document}